\documentclass[11pt]{article}

\usepackage{cyr}
\usepackage{epsf}
\usepackage[cp1251]{inputenc}
\usepackage[russian]{babel}
\usepackage{amssymb}
\usepackage{graphicx}
\usepackage{textcomp}
\usepackage{tipa}
\usepackage{lscape}

\def\*{$^{*}$}

\sloppypar

\begin{document}
\baselineskip=13pt

\begin{table}[t]
\begin{tabular}{l}
\scriptsize ISSN 1063-7737, Astronomy Letters, 2010, Vol. 36, No. 2, pp. 116-133. \copyright\ Pleiades Publishing, Inc., 2010.\\
\scriptsize Original Russian Text \copyright\ E.G. Chmyreva, G.M. Beskin, A.V. Biryukov, 2010, published in Pis'ma v Astronomicheskii Zhurnal, 2010,\\
\scriptsize Vol. 36, No. 2, pp. 125-143.\\
\hline\hline
\end{tabular}
\end{table}

\noindent  
\large{\bf SEARCH FOR PAIRS OF ISOLATED RADIO PULSARS - COMPONENTS IN DISRUPTED BINARY SYSTEMS}\\
\normalsize
\\
\noindent 
{\bf L. Chmyreva$^{1*}$, G.M. Beskin$^{2}$, A.V. Biryukov$^{1}$}\\
\\
\noindent 
{\it Sternberg Astronomical Institute, Universitetskii pr. 13, Moscow, 119992 Russia}$^1$\\ {\it Special Astrophysical Observatory, Russian Academy of Sciences, Nizhnii Arkhyz,
369167 Karachai-Cherkessian Republic, Russia}$^2$

\vspace{2mm}

\sloppypar 
\vspace{2mm}
\noindent 

{\bf Abstract}—We have developed a method for analyzing the kinematic association of isolated relativistic
objects—possible remnants of disrupted close binary systems. We investigate pairs of fairly young radio
pulsars with known proper motions and estimated distances (dispersion measures) that are spaced
no more than 2–3 kpc apart. Using a specified radial velocity distribution for these objects, we have
constructed 100–300 thousand trajectories of their possible motion in the Galactic gravitational field on
a time scale of several million years. The probabilities of their close encounters at epochs consistent with
the age of the younger pulsar in the pair are analyzed. When these probabilities exceed considerably their
reference values obtained by assuming a purely randomencounter between the pulsars under consideration,
we conclude that the objects may have been gravitationally bound in the past. As a result, we have detected
six pulsar pairs (J0543+2329/J0528+2200, J1453–6413/J1430–6623, J2354+6155/J2321+6024,
J1915+1009/J1909+1102, J1832–0827/J1836–1008, and J1917+1353/J1926+1648) that are companions
in disrupted binary systems with a high probability. Estimates of their kinematic ages and velocities at binary disruption and at the present epoch are provided.

\noindent{\bf DOI:} 10.1134/S1063773710020040

\vfill
\noindent\rule{8cm}{1pt}\\
{$^*$ \scriptsize E-mail $<$lisa.chmyreva@mail.ru$>$\normalsize}

\clearpage
\twocolumn

\section*{INTRODUCTION}
\noindent
Neutron stars (NSs) are among the fastest Galactic
objects. Their tangential velocities determined
from measured proper motions reach hundreds or
even thousands of kilometers per second (see, e.g.,
Hobbs et al. (2005) and references therein). NSs
probably receive such kicks during asymmetric supernova
explosions (Iben and Tutukov 1996; Helfand
and Tademaru 1977; Shklovsky 1970; Dewey and
Cordes 1987) or due to disruption by the explosions
of the binary systems whose components they were
(Blaauw 1961; Gott et al. 1970). The third possibility
is the birth of NSs during the explosions of highvelocity
massive stars that have been ejected from
young star clusters by dynamical processes in their
cores (Poveda et al. 1967; Gvaramadze 2007; Gvaramadze
et al. 2008, 2009).

At least 50\% of all stars are known to be members
of binary and multiple systems (Batten 1967; Zasov
and Postnov 2006; Duquennoy and Mayor 1991;
Halbwachs et al. 2003) and, accordingly, many of the
relatively young NSs that we observe as pulsars must
have been formed precisely in binary systems.

NSs are the final evolutionary stage of stars with
initial masses in the range from 8–10 to 25–30 $M_{\odot}$
(see, e.g., Postnov and Yungelson 2006; Bethe and
Brown 1998), although there is evidence that the
initial masses can be even higher (Muno et al. 2006).
Most such close binary systems (CBSs) are disrupted
by a supernova explosion, for which the loss of half of
itsmass is sufficient in the simplest case of a symmetric
explosion and a circular orbit; for an asymmetric
explosion, this fraction can be even smaller.

Nevertheless, when one of the stars in a massive
binary system explodes as a supernova at the end of its
life, this does not necessarily lead to binary disruption
— about 40\% of the pairs withstand the explosion
(Bethe and Brown 1998). Since the components of
a massive CBS exchange mass during its evolution,
only the C–O core might remain of the primary (more
massive) star at the time of its collapse. For this reason,
the mass lost during the first supernova explosion
may prove to be insufficient for binary disruption;
only the orbits of the components become eccentric
and the center of mass acquires a velocity of about
100 km s$^{-1}$. A CBS with a relativistic component, a
NS or a black hole (BH), emerges (see, e.g., Zasov
and Postnov 2006).

When the secondary component collapses during
the subsequent evolution (the second supernova explosion
occurs) and the second relativistic object is
formed, the pair is generally disrupted: the binary as a
whole is already less massive and, hence, the ejected
mass (higher than that in the first case) most often
exceeds the value needed for disruption (Zasov and
Postnov 2006).

The high velocities of pulsars serve as an additional
argument that many of them were members of
disrupted binary systems in the past. The slingshot
effect acts during the disruption of a binary system,
and this is one of the most natural mechanisms for
an increase in the space velocities of disruption remnants.
The kick received by the pulsar during an
asymmetric supernova explosion in conjunction with
the orbital angular momentum can accelerate the
NS significantly (Bailes 1989; Kiel and Hurley 2009).
However, this mechanism does not act at every CBS
disruption.

These currently isolated NSs may prove to be a
convenient tool for investigating the disruptionmechanism.
However, the kinematic characteristics of isolated
pulsars per se do not allow even the contribution
from additional velocities of a different nature to be
estimated. The situation will change radically if we
were to find two pulsars that used to be companions
in a disrupted CBS.

Considering pairs of fairly close pulsars with
known proper motions and distances and specifying
the distribution of their space velocities, we can trace
themotion of the objects in the past and determine the
distance of their closest approach. If its epoch turns
out to be consistent with the characteristic age $\tau_{ch}$
of one of the pulsars ($\tau_{ch}=P/2\dot P$, where $P$ is the
pulsar period), then this pair can be assumed to have
a common origin. The significance of this hypothesis
can be estimated by comparing the probability of the
encounter found with that for any two objects with
arbitrary proper motions. The presence of a supernova
remnant in the corresponding Galactic region that is
also consistent in age with one of the pulsars can
serve as an additional criterion for the validity of
the identification. Accepting the hypothesis about a
common origin will allow us to analyze the collapse
and second supernova explosion kinematics and the
CBS disruption dynamics as well as to ascertain the
relationship between the characteristic and actual
pulsar ages.

Obviously, it is possible to establish membership
in the same CBS only at disruption in the second
supernova explosion. The epoch of closest approach
must then coincide with the age of the younger pulsar.
If, however, the binary was disrupted after the
first explosion, then the second (intact) star evolves
independently and can recede to hundreds of parsecs
before it collapses $10^5-10^6$ yr later. The pulsar born in
this asymmetric explosion that received an additional
kick will change the direction of its motion, which
will make the detection of its companion virtually
impossible. Thus, the total number of disrupted pairs
can be larger.

Another scenario inaccessible to analysis by the
proposed method is that the massive progenitor stars
of pulsars in the past were ejected from young clusters
and, accordingly, acquired a high space velocity
(Gvaramadze 2007; Gvaramadze et al. 2008, 2009).
In that case, they evolve and collapse as isolated stars.
If the explosion does not change greatly the direction
of their velocities, then tracing the trajectories of the
two emerged young NSs into the past will only point
to the same cluster, i.e., the encounter will be fairly
close, but the stars cannot be said to have been gravitationally
bound in the past. Therefore, on the one
hand, the pairs of pulsars whose trajectories start near
young star clusters should be dealt with carefully (see,
e.g., Vlemmings et al. 2004) and, on the other hand,
it should be understood that the described situation
is unlikely. The progenitor stars ejected from clusters
must be massive enough to subsequently explode as
SNe II. Only about 20 fast massive stars are known
to date, and only two of them have a mass higher
than 8 $M_{\odot}$ (Przybilla et al. 2008; Heber et al. 2008).

The age difference between the components (i.e.,
the time elapsed between the first and second supernova
explosions) for massive binaries should not
exceed several million years. The latter requirement
follows from the characteristic lifetime of massive
main-sequence stars. If the time between the explosions
is short, then the characteristic ages of both
components can coincide. 

The question about the actual ages of pulsars
remains an open one. The characteristic age, particularly
for young pulsars, can serve only as a reference.
Faucher-Gigu\'ere and Kaspi (2006) analyzed
the relationship between the actual and characteristic
ages and showed that they could differ by a factor
of several. Additionally, $\tau_{ch}$ does not need to be an
upper limit for the actual age (see, e.g., Gaensler and
Frail 2000). The described approach to identifying
companion pulsars will help to estimate their ages
more accurately.

Isolated BHs that were previously members of
disrupted CBSs together with pulsars can also be
searched for by a similar method. The BH is the first
to be formed at the end of the evolution of the more
massive (> 30 $M_{\odot}$) star, following which the binary
survives with a probability of 40\% and is disrupted after
the second supernova explosionwith the formation
of a NS (Bethe and Brown 1998). In this case, since
the BH is massive, it will acquire a lower velocity and,
hence, will be not far from the place of CBS disruption
(the pulsar birthplace). Given the kinematic characteristics
of the pulsar (as one of the components of the
disrupted binary), the BH localization region where
it makes sense to search for it as a peculiar source
of radiation with various frequencies can be limited
in space (and in the sky) (Beskin and Karpov 2005).
The remnant of the second supernova near which the
pulsar trajectory passes if their ages are consistent,
can be an important reference in this search.

To localize the BH in such a way, we suggest using
pulsars with accurately measured proper motions
and distances. This accuracy will directly determine
the size of the region (from 1 to 10 square degrees)
where the BH is located under the assumption that
it exists. In these regions, the catalogs of starlike
sources in all ranges will be cross identified and peculiar
objects (primarily those without any lines in
their optical spectra) will be selected to search for
superfast variability in them, a basic property of an
isolated accreting BH (Beskin et al. 1997; Beskin and
Karpov 2005). Pulsars with young characteristic ages
will be used, where possible. In this case, the spatial
separation of the objects (the BH and the pulsar) is
definitely small (even given the inaccuracy of the age
estimate), which will facilitate the search for the BH.

Note separately that runaway OB stars can also be
used to search for NSs and BHs (see Prokhorov and
Popov (2002) and references therein).

In this work, we search for the pairs of isolated radio
pulsars that were bound in the past. The detection
of such pairs will allow one to understand the CBS
disruption dynamics and to estimate the velocities
acquired by the pulsars during a supernova explosion,
their radial velocities, ages, and type of explosion.

Several dozen pulsars associated with supernova
remnants are known to date. The most comprehensive
overview of such associations can be found in
Tian and Leahy (2004). These are mostly young pulsars
located either within or in the immediate vicinity
of their remnants. However, we are interested in the
cases where the pulsar is fast and old enough to have
left the neighborhood of the remnant (observed until
now).

The attempts to discover disrupted pulsar pairs
were previously made by Bisnovatyi-Kogan and
Komberg (1974), Gott et al. (1970), and
Wright (1979). The next step, an analysis of the
possible association of a pair of moderately young
pulsars spaced well apart for which the parallaxes and
proper motions are known with a good accuracy, was
undertaken by Vlemmings et al. (2004). Below, we
discuss this result.

In the next section, we describe in detail the
methodology and the assumptions made here. Next,
we present the results of our investigation of specific
pulsar pairs. The work is completed by a discussion
of the results obtained.

\section*{THE METHOD}
\subsection*{Formulation of the Problem}
\noindent
Searching for the kinematic association of two
pulsars with known proper motions consists in determining
the closest approaches of their reconstructed
trajectories in the Galaxy in the past. Obviously, the
pulsars with a common origin (i.e., those that were
the components of the same binary system) were
closest to each other at the time of CBS disruption.
This epoch corresponds to the current age of the
younger pulsar (recall that we analyze the case where
the system was disrupted after the second supernova
explosion).

We consider all motions in a rectangular coordinate
system whose origin is placed at the Galactic
center. One of its axes is parallel to the direction of
the Sun, the second axis is directed along the velocity
of the local standard of rest, and the third axis is
perpendicular to the Galactic plane and complements
the first two to a right-handed triple of vectors. In
such a coordinate system, the radius vector of the
Sun is ${\bf r}_{\odot} = \{-8.5$ $kpc$ $, 0, 0\}$ (see, e.g., Mihalas and
Binney 1981).

The law of pulsar motion ${\bf r}(t)$ in the Galactic
gravitational potential $\varphi_G({\bf r})$ is the solution of the
equation of motion

\begin{equation}
\ddot {\bf r} = -\nabla\varphi_G({\bf r})
\label{eqn:motion}
\end{equation}
with the initial conditions

\begin{equation}
{\bf r}_0 = {\bf r}(t = 0), {\bf V}_0 = {\bf V}(t = 0),
\label{eqn:initial}
\end{equation}
\noindent
where the present epoch corresponds to the initial
time. For known functions ${\bf r}_{1,2}(t)$ for the pulsar pair,
the separation $\rho$ between them at the past time $-t$ is
$|{\bf r}_1(-t)- {\bf r}_2(-t)|$. Associating the pulsars, we seek
for such an epoch $T$ consistent with the age of the
younger pulsar (no older than its several characteristic
ages) that

\begin{equation}
|{\bf r}_1(-T) - {\bf r}_2(-T)| = \min\rho.
\end{equation}

A low value of $\min\rho$ can indicate that these objects
were gravitationally bound in the past.

\subsection*{Solution of the Equation of Motion}
\noindent
Since the Galactic gravitational potential is not
spherically symmetric (see below), the solution of
Eq. (1) is generally obtained numerically, while the
vectors ${\bf r}_0$ and ${\bf V}_0$ are determined directly from observations.
For the current radius vector of a pulsar
with Galactic coordinates $l$ and $b$ and a heliocentric
distance $d$, we have

\begin{equation}
{\bf r}_0 = d\{\cos b \cos l, \cos b \sin l, \sin b\} + {\bf r}_{\odot}.
\end{equation}
\\
Its current velocity vector ${\bf V}_0$ is specified by the
known proper motion $\{\mu_{l}\cos b, \mu_{b}\}$, distance $d$, radial
velocity $V_{r}$, and solar velocity vector

\begin{equation}
\dot {\bf r}_{\odot} = {\bf V}_{\odot, rot} + {\bf V}_{\odot, LSR}.
\end{equation}
\\
\noindent
Here, ${\bf V}_{\odot, rot} = \{0, V_{\odot, rot}, 0\}$ is the Galactic rotation
velocity of the local standard of rest (LSR). It is
determined by the Galactic rotation curve (i.e., by
the potential $\varphi_G({\bf r})$) and is about 220 km s$^{-1}$. The
vector ${\bf V}_{\odot,LSR}$ is the solar velocity relative to the LSR;
it is directed to the point with $l=53^{\circ}$ and $b=25^{\circ}$
(apex) and its magnitude is $16.5$ km s$^{-1}$ (Mihalas and
Binney 1981).

The proper motion of a pulsar specifies its transverse
velocity ${\bf V}_t$ whose magnitude (in km s$^{-1}$) is

\begin{equation}
V_{t} = 4.74 \cdot d \cdot \sqrt{ (\mu_l \cos b)^2 + {\mu_b}^2 },
\end{equation}
\noindent
where $\mu_l$ and $\mu_b$ are in [mas yr$^{-1}$] and $d$ is in kpc. The
pulsar radial velocity consists of a secular component
${\bf V}_{r,rot}$ attributable to the Galactic rotation of the pulsar
LSR (i.e., also to the Galactic rotation curve) and
a peculiar velocity ${\bf V}_{r,p}$. Thus,we finally obtain
\begin{equation}
{\bf V}_0 = {\bf V}_{r,rot} + {\bf V}_{r,p} + {\bf V}_t + \dot {\bf r}_{\odot}.
\end{equation}

\subsection*{The Models Used}
\noindent
Apart from the spherical coordinates, only the
proper motions of pulsars and, more rarely, their distances,
when the parallax $\pi$ has been measured (the
distances are usually calculated from the dispersion
measure), are generally accessible to us directly from
observations. All of these quantities are known with
a finite accuracy and the distances are most often
model-dependent.

The main problem in studying the kinematics of
pulsars is the absence of direct measurements of their
radial velocities. At the same time, the source of information
about the latter is the distribution function
of one (any) of the three peculiar velocity components
of a pulsar at the time of its birth (e.g., at the time of
binary disruption). The total peculiar velocity is assumed
to be isotropic relative to the LSR. Therefore,
the same function can also be used for the peculiar
component of the pulsar radial velocity. 

It is constructed from the set of data on all of the
investigated pulsars as a solution of the maximum
likelihood problem. At present, the question about
its unambiguous choice is sill an open one. Various
versions have been proposed: Gaussian and non-Gaussian, one- or two-component distributions (see,
e.g., Hobbs et al. 2005; Cordes and Chernoff 1998).
Faucher-Gigu\'ere and Kaspi (2006) compare in detail
the results available to date and conclude that it is so
far hard to unambiguously choose a particular model,
because all models are consistent with the observational
data.

Thus, in the modeling task (in contrast to theoretical
constructions), the choice of a particular distribution
is not so important.We used it in the form of a
sum of two Gaussians (Arzoumanian et al. 2002):

$$p(V_{r,p})=\frac{w_1}{\sqrt{2\pi}\sigma_1}\exp{\left(-\frac{V_{r,p}^2}{2\sigma_1^2}\right)}
\label{eqn:p(V_r,p1)}
$$
\begin{equation}
+\frac{1-w_1}{\sqrt{2\pi}\sigma_2}\exp{\left(-\frac{V_{r,p}^2}{2\sigma_2^2}\right)},
\label{eqn:p(V_r,p2)}
\end{equation}
\noindent
where $\sigma_1=90$ km s$^{-1}$, $\sigma_2=500$ km s$^{-1}$ and $w_1=0.4$.

As was said above, it describes the velocities of
the pulsars at the epoch of their birth and can change
by now. It is easy to estimate the variations $\Delta V$ in
velocity magnitudes as a result of the motion in the
Galactic potential:
\begin{equation}
\Delta V \sim \tau \ddot r,
\label{eqn:deltaV}
\end{equation}
\noindent
where $\tau$ is the pulsar age and $\ddot r$ is the magnitude of
its acceleration. Vlemmings et al. (2004) used the
characteristic age $\tau_{ch}$ as $\tau$ and the acceleration $\ddot z$ in a
direction perpendicular to the Galactic plane instead
of $\ddot r$. However, they investigated a pair of moderately
young pulsars with ages $\tau_{ch} \sim 3$\texttimes$10^6$ yr. In this time,
correction (9) may prove to be significant and, at the
same time, significantly different from the true one,
since $\tau_{ch}$ is not a good age estimate and because the
corresponding integral is replaced with product (9).

Although here we investigate pulsars with considerably
younger ages, to estimate the effect of the
Galactic gravitational potential on their motion, we
simulated sets of pulsar trajectories that began their
motion in the Galactic plane (where the effect will be
at a maximum), at Galactocentric distances of $2.5 - 10.5$ kpc (with a step of 2 kpc). We obtained changes
in peculiar velocity magnitudes and deviations of the
trajectories from rectilinearity after $10^5$, $5$\texttimes$10^5$, $10^6$ and $3$\texttimes$10^6$ yr.

Function (8) was used as the distribution of pulsar
initial peculiar velocity components. The gravitational
potential $\varphi_G({\bf r})$ was specified by an expression
from Carlberg and Innanen (1987) and Kuijken and
Gilmore (1989):
$$
\varphi_G(r,z)=-\frac{GM_{dh}}{\sqrt{(a_G+\sum_{i=1}^3\beta_i\sqrt{z^2+h_i^2})^2+b_{dh}^2+r^2}}
$$

\begin{equation}
-\frac{GM_b}{\sqrt{b_b^2+r^2}}-\frac{GM_n}{\sqrt{b_n^2+r^2}},
\end{equation}
\noindent
where $r$ is the distance from the Galactic center and
$z$ is the distance from the Galactic plane. This is
a three-component axisymmetric potential that includes
the contribution from the disk–halo, bulge,
and nucleus; its parameters are presented in Table 1.

\begin{table}[t]

{\small {\bf Table 1.} Parameters of the Galactic potential (Carlberg
and Innanen 1987; Kuijken and Gilmore 1989)\normalsize}\label{tab:gal}

\vspace{5mm}
\begin{tabular*}{3.5in}{l|l|l|l} \hline\hline
Const-&Disk-halo(dh)&Bulge(b)&Nucleus(n)\\
ant&&&\\ \hline
$M$&1.45\scriptsize\texttimes\normalsize$10^{11}$ $M_{\odot}$&9.3\scriptsize\texttimes\normalsize$10^9$ $M_{\odot}$&1.0\scriptsize\texttimes\normalsize$10^{10}$ $M_{\odot}$\\
$\beta_1$&0.4&&\\
$\beta_2$&0.5&&\\
$\beta_3$&0.1&&\\
$h_1$&0.325 kpc&&\\
$h_2$&0.090 kpc&&\\
$h_3$&0.125 kpc&&\\
$a_G$&2.4 kpc&&\\
$b$&5.5 kpc&0.25 kpc&1.5 kpc\\ \hline
\end{tabular*}
\end{table}

Obviously, the closer the pulsar to the Galactic
center, the more nonuniform and nonrectilinear its
motion. The limiting case for a Galactocentric distance
of 2.5 kpc is presented in Figs. 1 and 2, which
shows the changes in peculiar velocity magnitudes
for pulsars and the deviations of their trajectories
from a rectilinear shape. We see that their motion
is almost uniform (compared to the parameters of
distribution (8)) and rectilinear on time scales that
do not exceed $10^6$ yr. All distributions are bimodal,
because function (8) is a two-component one. 

For pulsars that began their motion from a Galactocentric
distance of 10.5 kpc, the scatter  $\sigma[\Delta V]$ does
not exceed 4 km s$^{-1}$ even for $T = 3$\texttimes$10^6$ yr and the
mean $\mu[\Delta r] = 21$ pc.

Thus, when the motion of pulsars in the past is
investigated on time scales of $\sim 10^6$ yr or shorter, including
the effect of $\varphi_G({\bf r})$ in the form of correction (9)
seems superfluous and can even lead to additional
errors in the results, since the actual age $\tau$ can differ
from the characteristic one $\tau_{ch}$ by a factor of several
(see, e.g., Faucher-Gigu\`ere and Kaspi 2006). In contrast,
distribution (8) also describes well the current
pulsar velocities.

The initial conditions ${\bf r}_0$ and ${\bf V}_0$ are specified in the
form of distributions that reflect the inaccuracies in
the distances and velocities. 

\subsection*{Randomization of Initial Conditions}
\noindent
As has been noted above, all of the pulsar characteristics
known from observations are specified with
a finite accuracy. As in Vlemmings et al. (2004), we
assume the pulsar proper motions to be distributed
according to normal laws $N(\mu_l, \sigma^2_{\mu_l})$ and $N(\mu_b, \sigma^2_{\mu_b})$
with means equal to the measured values and variances
equal to the squares of themeasurement errors.
A similar assumption is also used for the distances
(mean $d$ and variance $\sigma^2_d$). Here, we estimated them
from highly accurate measurements of the dispersion
measure (DM) based on the NE2001 model of the
Galactic free electron density distribution (Cordes
and Lazio 2002) with a relative error of about 30\%.

The pulsar equatorial coordinates are known
with very high accuracies of fractions of an arcsecond,
which gives an uncertainty in the position of
$10^{-3} - 10^{-2}$ pc at distances to the objects of several
kpc. These values are considerably lower than the
distance errors $\sigma_d$ and, hence, we assumed $\alpha$ and $\delta$
(as well as the calculated $l$ and $b$) to be known exactly.

Since, as was shown above, the pulsar motion on
time scales of $10^6$ yr remains almost uniform, the
uncertainty in the calculated ${\bf r}(t)$ will generally be the
same as that for the distance $d$.

In turn, $\sigma_{\mu_l}$ and $\sigma_{\mu_b}$, along with the error in the distance
$d$, will lead to randomization of the magnitude
(and direction) of the current velocity ${\bf V}_0$. For $d \sim 2 - 4$ kpc, proper motions of $\sim 10$ mas yr$^{-1}$, and their
relative errors of $\sim 30\%$, the error in the transverse
velocity will $\sim 100$ km s$^{-1}$, which will subsequently,
on time scales up to $10^6$ yr, lead to an error in the
position ${\bf r}(t)$ of 100 pc. This is much smaller than
the contribution from the inaccuracy in the initial
distance $d$.

\subsection*{Simulations}
\noindent
It is clear from the aforesaid how the distributions
$p_r({\bf r}_0)$ and $p_v({\bf V}_0)$ of initial vectors ${\bf r}_0$ and ${\bf V}_0$ in (2)
are obtained. Searching for the closest encounters
of two pulsars is reduced to constructing the distribution
of the probability $P(\rho)$ that the minimum
separation between them (on a given time scale in the
past) does not exceed $\rho$.

This distribution was constructed by the Monte
Carlo method; solving Eq. (1) and simulating the sets
of pairs of pulsar trajectories with various realizations
of ${\bf r}_0$ and ${\bf V}_0$ from $p_r({\bf r}_0)$ and $p_v({\bf V}_0)$, we traced them
into the past to a time of 1–2 Myr (exceeding the
characteristic age of the younger pulsar) and calculated
the minimum separation $\rho_{min}$ between them.
Since the effect of the Galactic gravitational potential
(10) in our problem is negligible, we restricted
ourselves to several points with a step of 500 000 yr
determined by the Runge–Kutta–Feldberg method
for each trajectory. The pulsar motion between them
was assumed to be uniform and rectilinear, which
allowed the speed of our simulations to be increased
significantly.

Together with the distribution $P(\rho)$, we obtained
the distribution $T(\rho)$ for the epochs of closest approaches.
Next, after comparison of the former distribution
with the reference one $P_0(\rho)$, we reached the
conclusion about whether the kinematic association
of the pulsar pair under study was possible.

\subsection*{Reference Probability}
\noindent
Once the distribution $P(\rho)$ has been obtained, we
need a criterion that would allow us to determine
how significant the probability that a given pair of
pulsars approached to some small distance $\tilde \rho$ in the
past is. If the probability $P(\tilde \rho)$ is too high for the
simulated number of approaches to a distance $\leqslant \tilde \rho$ to
be explained only by random factors, then the
\onecolumn
\begin{figure}[t]
\centering
\includegraphics[width=13cm]{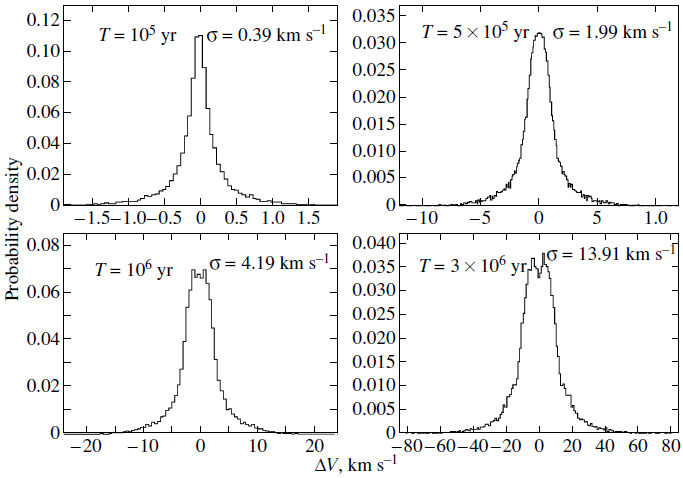}\\
\small {\bf Fig.1:} Distribution of changes in peculiar velocity magnitudes for pulsars under the Galactic gravitational field that began
their motion at a Galactocentric distance of 2.5 kpc.\normalsize
\label{fig:v25}
\end{figure}

\begin{figure}[b]
\centering
\includegraphics[width=13cm]{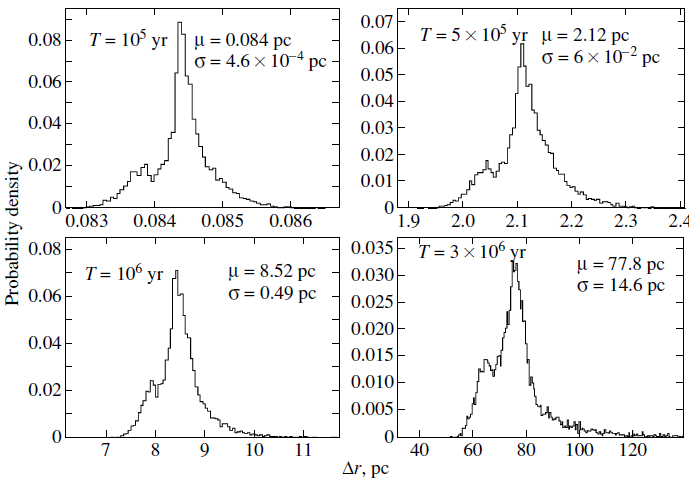}\\
\small {\bf Fig.2:} Distribution of changes in final coordinates for pulsars under the Galactic gravitational field that began their motion at
a Galactocentric distance of 2.5 kpc.\normalsize
\label{fig:r25}
\end{figure}
\clearpage
\twocolumn
\noindent
pair
under study proves to be kinematically singled out
among the others. There is reason to believe that
these pulsars were once actually close to each other
and to suggest that they constituted a single disrupted
binary system.

This necessary criterion consists in comparing the
probabilities $P(\tilde \rho)$ and $P_0(\tilde \rho)$, where the latter is the
probability for two randomly chosen pulsars to
be at a distance no larger than $\tilde \rho$ during their motion.
Obviously, this probability should be calculated by
taking into account the pulsar density in the Galaxy.

Since there are pulsars of different ages and different
origins in the Galaxy, $P_0(\tilde \rho)$ is equivalent to
the probability of detecting two (or more) pulsars
at a distance $\leqslant \tilde \rho$ when a given region of space is
observed. Clearly, the larger this region and the higher
the pulsar density in it, the higher the probability of
such a detection.

Thus, to calculate $P_0(\tilde \rho)$, we must specify, first,
the pulsar density distribution over the Galaxy and,
second, the sizes and localization of the region of
space of interest. Here, we used the radial pulsar
surface density distribution derived by Yusifov and
K\"{u}c\"{u}k (2004):
$$\Sigma(R)=A\left(\frac{R + R_1}{R_\odot + R_1}\right)^a\times$$
\begin{equation} 
\exp{\left[-b\left(\frac{R-R_\odot}{R_\odot + R_1}\right)\right]},
\label{eqn:sigma(R)}
\end{equation}
\noindent
where $A = 37.6 \pm 1.9$ kpc$^{-2}$, $a = 1.64 \pm 0.11$, $b = 4.01 \pm 0.24$, $R_1 = 0.55 \pm 0.10$ kpc, and $R_\odot$ = 8.5 kpc,
and a similar distribution in a direction perpendicular
to the Galactic plane, whose choice and the optimality
of the parameter $\langle z \rangle = 50$ pc are justified by Faucher-Gigu\`ere and Kaspi (2006):

\begin{equation}
p(z)=\frac{1}{2\langle z \rangle}\exp\left(-\frac{|z|}{\langle z\rangle}\right).
\label{eqn:p(z)}
\end{equation}

Collectively, these distributions specify the average
number of pulsars $\eta(\bf r)$ that can actually be detected
in a Galactic region with coordinates of its
center ${\bf r} = \{x, y, z\}$ and sizes $\tilde \rho$ [pc]:

\begin{equation}
\eta({\bf r}) = 10^{-6} \tilde \rho^2 \Sigma(\sqrt{x^2 + y^2}) \int_{z -\tilde \rho/2}^{z + \tilde
\rho/2}{p(z)dz}.
\end{equation}
\noindent
Here, $x,y,z$ are the rectangular coordinates in the
coordinate system associated with the Galaxy and described
above. The distribution of pulsars is assumed
to be stable for several million years.

Calculating the probability $P(\tilde \rho)$ from our simulations,
we automatically obtain the region (${\bf r}_{\tilde \rho }$, ${\bf r}_{\tilde \rho }+d{\bf r}_{\tilde \rho }$) where the pulsars approach each other to distances
no larger than $\tilde{\rho}$. When $\tilde{\rho} \ll dr_{\tilde \rho}$, the number of
pulsars in the volume $\tilde \rho^3$ obeys a Poisson distribution.
Therefore, we divided the entire region into nonoverlapping
cells ${\bf r}_k$ ($k = 1..N$) with sizes $\tilde \rho$ in each
of which the probability of detecting fewer than two
pulsars is specified by the expression

\begin{equation}
\xi_k = \sum_{n < 2}{\frac{[\eta({\bf r}_k)]^n}{n!} \exp[-\eta({\bf r}_k)]}.
\end{equation}

Finally, $P_0(\tilde \rho)$, the probability of detecting two or
more pulsars at a distance $\leqslant \tilde \rho$ somewhere in the
Galactic region (${\bf r}_{\tilde \rho }$, ${\bf r}_{\tilde \rho } + d{\bf r}_{\tilde \rho }$), was calculated from
the formula

\begin{equation}
P_0(\tilde \rho) = 1 - \prod_{k = 1}^{N}{\xi_k}.
\end{equation}

This quantity is indicative of whether a particular
pulsar pair stands out kinematically. For many pairs,
$P(\rho)$ and $P_0(\rho)$ were comparable (see below), but
they differed significantly for some of them. In our
problem, the kinematic association of pulsars is based
on Monte Carlo calculations of the probability that
a specific pair of pulsars approached to a given (or
smaller) distance in the past. If this probability for
sufficiently close approaches (of the order of several
parsecs) is high compared to the reference one
$P(\rho) \gg P_0(\rho)$, then it is concluded that they have a
common origin.

\begin{figure}[h]
\includegraphics[width=9cm]{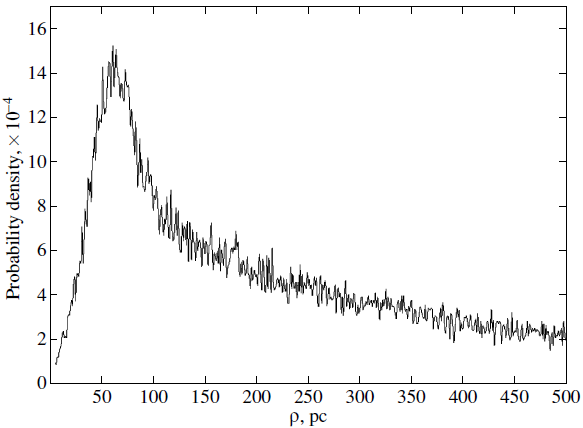}\\
\small {\bf Fig.3:} Distribution of minimum separations between the pulsars J1453–6413 and J1430–6623 derived from our simulations
of 15 000 trajectories.\normalsize
\label{fig:prob-den}
\end{figure}

\subsection*{Mutual Correspondence between the Adopted
Models}
\noindent
Here, we assume the distance to the Galactic
center to be $R_{\odot}$ = 8.5 kpc. The gravitational potential
(10) suggests the same $R_{\odot}$. It is based on
a standard Galactic rotation curve. In contrast, the
distribution of pulsar peculiar velocities obtained by
Arzoumanian et al. (2002) uses the TC93 models
of the Galactic free electron density (Taylor and
Cordes 1993), while we use NE2001. Their spatial
distribution is assumed to be uniform and the
gravitational potential is based on the papers by
Paczynski (1990) and Caldwell and Ostriker (1981),
where $R_{\odot}$ = 8 kpc. In general, as is pointed out in
the ``Models Used'' Section, the observational data
available to date do not allow one to unambiguously
choose a particular distribution of pulsar velocities.
We are talking not only about the specific parameters
but also about the shape of the distribution (for a
review, see Faucher-Gigu\`ere and Kaspi (2006)). All
of the possible versions do not differ from one another
dramatically. For our results to change qualitatively,
it is necessary that the simulated probabilities of close
encounters be several orders of magnitude higher
than the values obtained (as will be seen from the
results). Varying the shape of the velocity distribution
cannot lead to such a situation, especially since the
pulsar motion is investigated on time scales of several
hundred thousand years, when reasonable differences
in the distributions do not yet manifest themselves.
Therefore, the distribution of peculiar velocities from
Arzoumanian et al. (2002) is quite suitable for this
work.
\section*{RESULTS}
\subsection*{Calculations}
\noindent
The possibility of a kinematic association of pulsars
was considered for pairs that satisfied the following
conditions:

	(1) the proper motions ($ \mu_\alpha$, $\mu_\delta$) and distances
(parallax $\pi$, DM) are known;\\\indent
	(2) the current separation does not exceed 2–3 kpc;\\\indent
	(3) the characteristic age $\tau_{ch}=P/2\dot P$ of the
younger pulsar does not exceed $10^6$ yr (this is because
the binary disruption region is determined more
accurately for a small displacement of the pulsar from
its current position backward along the trajectory,
especially since we can be in error by a factor of
several while estimating the disruption time from the
characteristic age);\\\indent
	(4) the pulsar age difference does not exceed several
million years (this follows from the characteristic
lifetime of massive main-sequence stars). If the true
pulsar age is no more than $\sim$100 000 yr, then the supernova
remnant associated with it can survive (and
be detected) to the present day.

Here, we initially formed a sample of $\sim$200 isolated
radio pulsars with known proper motions and distances.
Their parameters were taken mostly from the
ATNF database (Manchester et al. 2005;
http://www.atnf.csiro.au/research/pulsar/psrcat/expert.html). We selected 16 pairs of objects from
them that met the above criteria. For each of these
pairs, we first simulated 10-15 thousand trajectories
(per each pulsar). Next, we selected 7 from the
16 pairs for which the probability density $p(\rho)=\frac{d}{d\rho}P(\rho)$ of the distribution of separations between the
components had a distinct sharp peak at $\rho\leqslant 500$ pc.

\onecolumn
\begin{table}[t]
\centering
{\small{\bf Table 2.} Pulsar parameters.\normalsize}\label{tab:data}
\vspace{5mm}

\begin{tabular}{c|c|c|c|c|c|c|c} \hline\hline
Pulsars & $\alpha$ & $\delta$ & D & $\mu_{\alpha}$ & $\mu_{\delta}$ & $\tau_{ch}$ & $\Delta$ \\
         & (hms) & (dms) & (kpc) & (mas yr$^{-1}$)& (mas yr$^{-1}$) & ($10^3$ yr) & (kpc) \\ \hline\hline
J0543+2329 & 05 43 09.660 & 23 29 05.00 & 2.06 & 19(7) & 12(8) & 253 & 0.46 \\
J0528+2200 & 05 28 52.308 & 22 00 01.00 & 1.61 & -20(19) &  7(9) & 1480 & \\ \hline
J1453-6413 & 14 53 32.737 & -64 13 15.59 & 2.08 & -16(1) & -21.3(8) & 1040 & 1.08 \\
J1430-6623 & 14 30 40.872 & -66 23 05.04 & 1.00 & -31(5) & -21(3) & 4490 & \\
\hline
J2354+6155 & 23 54 04.724 & 61 55 46.79 & 3.43 & 22(3) & 6(2) & 920 & 0.46 \\
J2321+6024 & 23 21 55.213 & 60 24 30.71 & 3.03 & -17(22) & -7(19) & 5080 & \\
\hline
J1915+1009 & 19 15 29.98290 & 10 09 43.780 & 6.27 & 4(7) & -10(13) & 420 & 2.09 \\
J1909+1102 & 19 09 48.69380 & 11 02 03.350 & 4.18 & -6(4) & 7(8) & 1700 & \\
\hline
J1832-0827 & 18 32 37.0200 & -08 27 03.64 & 4.85 &  -4(4) & 20(15) & 161 & 0.42 \\
J1836-1008 & 18 36 53.925 & -10 08 08.3 & 4.46 & 18(65) & 12(220) & 756 & \\
\hline
J1833-0827 & 18 33 40.3000 & -08 27 31.25 & 4.66 & -38(26) & -72(55) & 147 & 0.24 \\
J1836-1008 & 18 36 53.925 & -10 08 08.3 & 4.46 & 18(65) & 12(220) & 756 & \\
\hline
J1917+1353 & 19 17 39.7902 & 13 53 56.95 & 3.99 & 0(12) & -6(15) & 428 & 1.87 \\
J1926+1648 & 19 26 45.322 & 16 48 32.78 & 5.83 & 13(11) & -14(18) & 511 & \\ \hline

\end{tabular}
\end{table}

\begin{table}[b]

\vspace{-20mm}
\centering
{\small {\bf Table 3.} Pairs of pulsars that approached each other most closely in the past.\normalsize}\label{tab:result}

\vspace{5mm}
\begin{tabular}{p{1.9cm}|p{1cm}|p{0.5cm}|p{0.7cm}|p{0.5cm}|p{1.8cm}|p{1.6cm}|p{1.4cm}|p{1.3cm}|p{1.55cm}|p{1.5cm}} \hline\hline
Pulsars & N & $\rho_0$ & $\rho_{min}$ & $n(5)$ & $P(5)$ & $P_0(5)$ & $P_0(\rho_0)$ & $T(5)$ & $V_r(5)$ & $V(5)$ \\
         & & (pc) & (pc) & & & & & ($10^3$ yr) & (km s$^{-1}$) & (km s$^{-1}$) \\ \hline\hline
J0543+2329 & 182634 & 285 & 0.535 & 81 & 4.435\scriptsize\texttimes\normalsize 10$^{-4}$ & \textless 5\scriptsize\texttimes\normalsize 10$^{-10}$ & 0.249 & 325\scriptsize$\pm$\normalsize 92 & 50\scriptsize$\pm$\normalsize 131 & 322\scriptsize$\pm$\normalsize 85\\
J0528+2200 & & & & & & & & & 51\scriptsize$\pm$\normalsize 257 & 369\scriptsize$\pm$\normalsize 178\\
\hline
J1453-6413 & 199272 & 176 & 1.096 & 35 & 1.756\scriptsize\texttimes\normalsize 10$^{-4}$ & 3.1\scriptsize\texttimes\normalsize 10$^{-9}$ & 0.137 &  566\scriptsize$\pm$\normalsize 147 & -156\scriptsize$\pm$\normalsize 257 & 321\scriptsize$\pm$\normalsize 93\\
J1430-6623 & & & & & & & & & 319\scriptsize$\pm$\normalsize 334 & 401\scriptsize$\pm$\normalsize 118\\
\hline
J2354+6155 & 93978 & 182 & 1.79 & 8 & 8.513\scriptsize\texttimes\normalsize 10$^{-5}$ & 1.8\scriptsize\texttimes\normalsize 10$^{-9}$ & 0.094 &  279\scriptsize$\pm$\normalsize 50 & -253\scriptsize$\pm$\normalsize 185 & 609\scriptsize$\pm$\normalsize 102\\
J2321+6024 & & & & & & & & & -283\scriptsize$\pm$\normalsize 844 & 930\scriptsize$\pm$\normalsize 124\\
\hline
J1915+1009 & 222007 & 81 & 2.646 & 10 & 4.504\scriptsize\texttimes\normalsize 10$^{-5}$ & 2.02\scriptsize\texttimes\normalsize 10$^{-8}$ & 0.012 & 303\scriptsize$\pm$\normalsize 64 & -224\scriptsize$\pm$\normalsize 306 & 616\scriptsize$\pm$\normalsize 197\\
J1909+1102 & & & & & & & & & -28\scriptsize$\pm$\normalsize 59 & 283\scriptsize$\pm$\normalsize 4\\
\hline
J1832-0827 & 343000 & 55 & 3.098 & 3 & 8.746\scriptsize\texttimes\normalsize 10$^{-6}$ & 1.23\scriptsize\texttimes\normalsize 10$^{-8}$ & 0.00178 & 42\scriptsize$\pm$\normalsize 17 & 20\scriptsize$\pm$\normalsize 17 & 735\scriptsize$\pm$\normalsize 47\\
J1836-1008 & & & & & & & & & 4\scriptsize$\pm$\normalsize 5 & 2068\scriptsize$\pm$\normalsize 630\\
\hline
J1833-0827 & 239034 & 46 & 1.73 & 2 & 8.367\scriptsize\texttimes\normalsize 10$^{-6}$ & 2.5\scriptsize\texttimes\normalsize 10$^{-9}$ & 0.000753 & 26\scriptsize$\pm$\normalsize 1 & -360\scriptsize$\pm$\normalsize 360 & 1179\scriptsize$\pm$\normalsize 805\\
J1836-1008 & & & & & & & & & 226\scriptsize$\pm$\normalsize 157 & 4833\scriptsize$\pm$\normalsize 1034\\
\hline
J1917+1353 & 221335 & 42 & 3.399 & 2 & 9.036\scriptsize\texttimes\normalsize 10$^{-6}$ & \textless 5\scriptsize\texttimes\normalsize 10$^{-9}$ & 0.000593 & 696\scriptsize$\pm$\normalsize 186 & 491\scriptsize$\pm$\normalsize 192 & 712\scriptsize$\pm$\normalsize 346\\
J1926+1648 & & & & & & & & & -117\scriptsize$\pm$\normalsize 87 & 402\scriptsize$\pm$\normalsize 215\\ \hline

\end{tabular}
\end{table}
\twocolumn
\begin{figure}[t]
\hspace{-1cm}
\includegraphics[width=9.7cm]{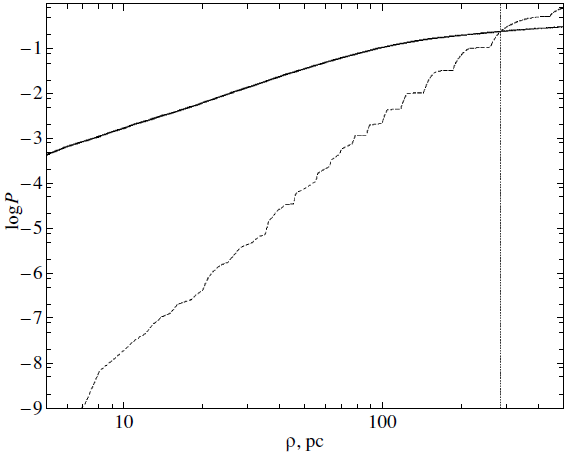}\\
\small {\bf Fig.4:} Pulsars J0543+2329 and J0528+2200. The probability of detecting the pulsar pair in the past at a distance of less
than $\rho$ = 5 pc. The solid and dashed curves represent, respectively, the probability $P$ determined during our simulations and
the probability $P_0$ of a chance (Poissonian) detection of two (or more) pulsars in a cell of space with a size $\rho$ based on the
distribution of their density in the Galaxy. The nonmonotonic behavior of the function $P_0(\rho)$ at $\rho$\textgreater 30 pc is explained by
computational noise. The vertical line marks $\rho_0$ = 285 pc.\normalsize
\label{fig:y19o17-1}
\end{figure}
\noindent
Figure 3 shows an example of such a distribution for
the pair of J1453–6413 and J1430–6623. Table 2
presents parameters of the seven pairs: the pulsar coordinates,
the distance to the pulsar (determined from
the DM), the proper motion, the characteristic age,
and the separation between the pulsars at the present
epoch. For each of them, we then simulated 100–300 thousand pairs of trajectories, constructed
the distributions $P(\rho)$, $P_0(\rho)$, and $T(\rho)$, and determined
the regions of their closest approaches in space (and
in the sky), starting from $\rho$ = 5pc.

For each of the seven pairs, the probabilities $P_0(\rho)$
at small $\rho$ ($\leqslant$50–300 pc) are considerably lower
than $P(\rho)$; for very small $\rho$, these probabilities differ
by orders of magnitude, which virtually rules out the
possibility of a purely chance detection of such close
encounters in simulations.

The results of our simulations are presented in
Table 3. For each pulsar pair, it gives: (1) the total
number N of simulated trajectory pairs; (2) the
distance $\rho_0$ starting from which the probability of a
random encounter is higher than that of a nonrandom
one:
$P_0(\rho$\textgreater$\rho_0)$\textgreater$P(\rho$\textgreater$\rho_0)$; (3) the minimum separation
found among the simulated trajectories $\rho_{min}$;
(4) the number n of trajectory pairs that approach to a
distance $\rho\leqslant$ 5 pc; (5) the corresponding probabilities
$P$ ($\rho\leqslant$ 5 pc) and $P_0$ ($\rho\leqslant$ 5 pc); (6) the probability
$P_0(\rho_0)$ of a random encounter at the point of intersection
between the distributions (in the volume
of size  $\rho_0$); (7) the median epoch $T$ and the 50\%
confidence interval that characterizes the approaches
to $\rho\leqslant$ 5 pc; (8) the median values and 50\% confidence
intervals of the current radial velocities $V_r$ for this
case; (9) the median values of the pulsar peculiar
velocities $V(5)$ at the epoch of close approach and the
50\% confidence interval. 

\subsection*{Discussion of Simulation Results}
\noindent
While analyzing the trajectories of the pulsar pairs
that may have a common origin, it is interesting (and
important) to analyze the velocity vectors of their
separation at CBS disruption, i.e., the magnitudes of
these velocities and the angles between them, first, to
test the theories of CBS disruption by a supernova
explosion and, second, to additionally verify the possibility
that the pulsars were bound (or unbound) in the
past. However, the quantities that can be obtained in
terms of a statistical analysis have their peculiarities.
Specifically, each individual value of the velocity or
angle carries little information. In our case, these
are only individual realizations of a random variable
that per se do not characterize the pulsar pair in
any way. Only the characteristics of their ensemble
are meaningful. In our case, these are the confidence
intervals, which are wide for these quantities. This
is understandable: the uncertainty in the distances
and radial velocities of the sample pulsars is very
high. Thus, for example, for the closest encounters of
the pair J0543+2329/J0528+2200 (at $\rho\leqslant$ 5 pc), the
magnitudes of their velocities ``immediately after the
separation'' lie with a 50\% probability in the intervals
from $\sim$230 to 400 km s$^{-1}$ and from 190 to 550 km s$^{-1}$,
respectively, while the angles between the velocities
lie in the interval from $\sim$40$^{\circ}$ to $\sim$120$^{\circ}$. Therefore, it
turns out to be difficult to reliably estimate the velocity
vectors at the time of the presumed separation, and
such an analysis was not performed in this paper.
Parallax measurements for the pulsars under consideration
would help greatly to solve this problem.

\subsubsection*{J0543+2329 and J0528+2200}
\noindent
The best results
were obtained for the pair of pulsars J0543+2329 and
J0528+2200 in our simulations of 182634 trajectory
pairs. The approach to 5 pc (or smaller) is realized for
81 trajectory pairs, which corresponds to the probability
$P=4.435$\texttimes$10^{-4}$ at a probability of a random
encounter $P_0$ for the same distances of less than
$5$\texttimes$10^{-10}$. The difference by six orders of magnitude
suggests that such a close encounter of the pulsars is
not random and is indicative of their common origin.
The value of $\rho_0$ for this pair is 285 pc. For clarity,
Fig. 4 shows the probability of detecting the pulsar
pair in the past at a distance smaller than $\rho$. (Since
the dependence is similar in shape for all succeeding
pulsar pairs, we omit similar figures for them.)
The smallest separation found between the pulsars in
the past is 0.535 pc. The epoch of close encounters
corresponds to $T\sim$ 300 000 yr ago, and the characteristic
age $\tau_{ch}$ of the younger pulsar J0543+2329 is
253 000 yr, which is in satisfactory agreement. If the
separation between the pulsars is assumed to be at a
minimum precisely at the birth time of J0543+2329,
then its actual age is close to 300 000 yr. Since the
older pulsar J0528+2200 has a characteristic age
of about 1.5 Myr, its progenitor had sufficient time
before its collapse, in agreement with the evolutionary
scenario for massive CBSs. The median means of the
current pulsar radial velocities (for approaches to a
distance smaller than 5 pc) are 50 and 51 km s$^{-1}$
for J0543+2329 and J0528+2200, respectively.
The
total space velocities of the pulsars relative to the LSR
at the time of binary disruption (when the pulsars
approached to $\rho\leqslant$ 5 pc) are 322 and 369 km s$^{-1}$.
Figure 5 shows the pulsar trajectories for which the
spatial approaches are closest.

\subsubsection*{J1453–6413 and J1430–6623}
\noindent
High probabilities
of close approaches are also characteristic of
the pair of J1453-6413 and J1430-6623. Out of the
199272 simulated trajectories, 35 pairs approached
to a distance $\rho\leqslant$ 5 pc, i.e., the probability $P(\rho\leqslant 5$ pc) is $1.756$\texttimes$10^{-4}$ and the probability of a random
encounter $P_0$ is approximately five orders of magnitude
lower: $3.1$\texttimes$10^{-9}$. The minimum separation
during the approach was 1.096 pc ($\rho_0$ for this pair is
176 pc) (see Fig. 6). The closest approaches occurred
about 570 000 yr ago. The characteristic age of the
younger pulsar J1453.6413 is about $10^6$ yr; given
the errors, this is a good correspondence, as in the
previous case.
The velocities at the binary disruption
(when the pulsars approached in the past) were about
300 km s$^{-1}$ for J1453-6413 and 400 km s$^{-1}$ for
J1430-6623. There is the OB association Pis 20
(Mel\textquoteright nik and Efremov 1995) inside the spatial region
of approaches to $\rho\leqslant$ 10 pc, an argument for the
pulsar birth and the binary disruption nearby (see,
e.g., Bobylev 2008). The center of the association is
located $\sim$80 pc away from the place in which the
approach of the pulsars is closest. In this case, the
motion of the association in the Galaxy is disregarded,
since the OB association with a velocity much lower
than that of the pulsar was displaced insignificantly
on time scales of several hundred thousand years.

\subsubsection*{J2354+6155 and J2321+6024}
\noindent
For the pair of
J2354+6155 and J2321+6024 (Fig. 7), the minimum
separation of the approach is 1.79 pc at its probability
$P=8.513$\texttimes$10^{-5}$. The probability of a random encounter
$P_0$ is approximately four orders of magnitude
lower; $\rho_0$ for this pair is also fairly large: 182 pc. The
approaches occurred about 280000 yr ago, and the
characteristic age of the younger pulsar J2354+6155
is 920000 yr - this difference is also acceptable.
\onecolumn

\begin{figure}[t]
\centering
\includegraphics[width=12cm]{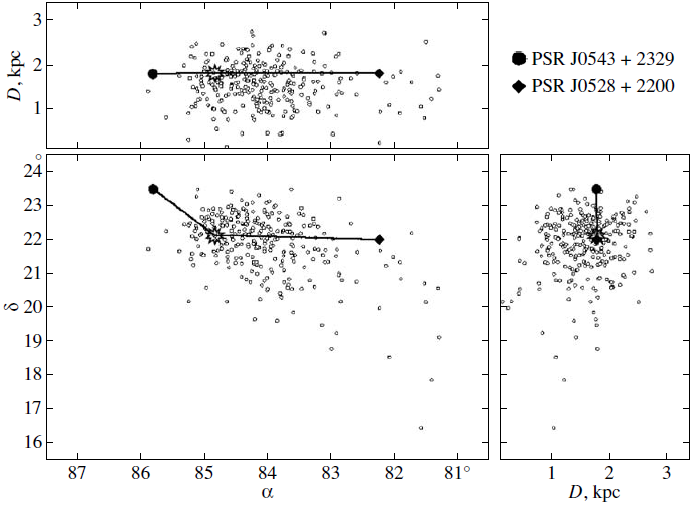}\\
\small {\bf Fig.5:} Pulsars J0543+2329 and J0528+2200. Projections of the spatial region in which the measured minimum separation
between the pulsars does not exceed $\rho$ = 10 pc. Each point on the plot is the place of closest approach of the pulsars for a
specific pair of trajectories. The circle and the diamond mark the position of the younger pulsar at the present epoch and the
position of the older pulsar, respectively. The lines indicate the trajectories at which the separation found between the pulsars
$\rho_{min}$ is smallest. The asterisk marks the region where it is reached. Here, $\rho_{min}$ = 0.535 pc.\normalsize
\label{fig:y19o17-2}
\end{figure}

\begin{figure}[b]
\centering
\includegraphics[width=12cm]{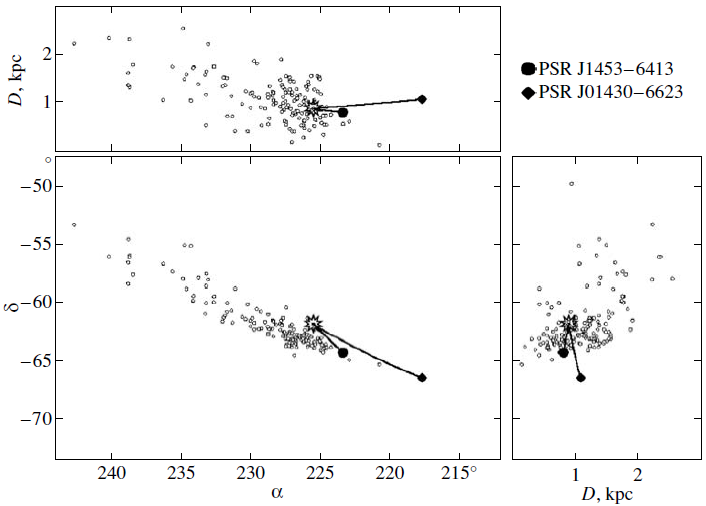}\\
\small {\bf Fig.6:} Same as Fig. 5 for the pulsars J1453–6413 and J1430–6623. The trajectories at which the separation found between
the pulsars, $\rho_{min}$ = 1.096 pc, is smallest. The points mark the places of the closest approach of specific trajectory pairs the
minimum separation between which is $\rho\leqslant$ 10 pc.\normalsize
\label{fig:o13o52-2}
\end{figure}

\begin{figure}[t]
\centering
\includegraphics[width=12cm]{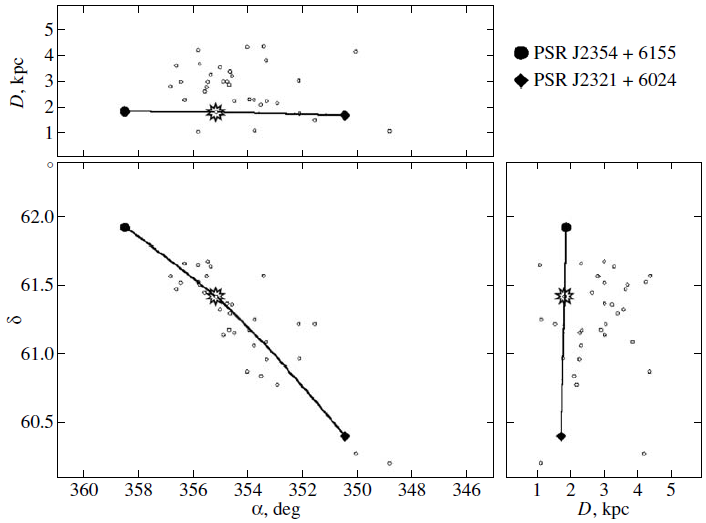}\\
\small {\bf Fig.7:} Same as Fig. 5 for the pulsars J2354+6155 and J2321+6024. The trajectories at which the separation found between
the pulsars, $\rho_{min}$ = 1.79 pc, is indicated by the lines. The points mark the places of the closest approach of specific trajectory
pairs the minimum separation between which is $\rho\leqslant$ 10 pc.\normalsize
\label{fig:o9o59-2}
\end{figure}

\begin{figure}[b]
\centering
\includegraphics[width=12cm]{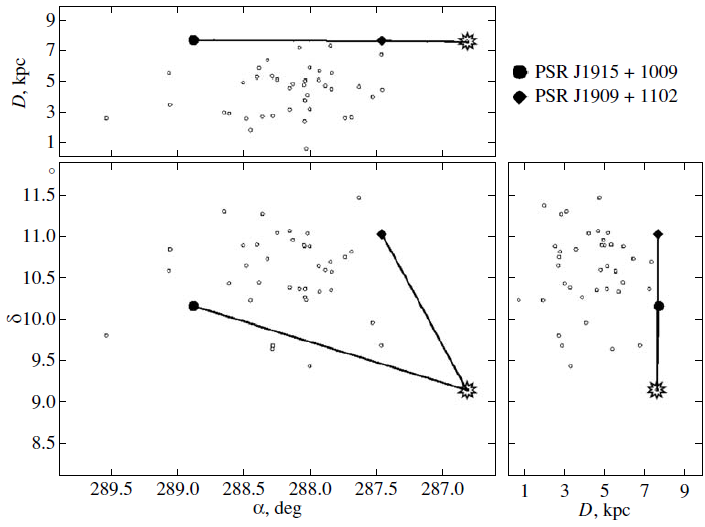}\\
\small {\bf Fig.8:} Same as Fig. 5 for the pulsars J1915+1009 and J1909+1102. The lines indicate the trajectories at which the separation
found between the pulsars, $\rho_{min}$ = 2.646 pc, is smallest. The pointsmark the places of the closest approach of specific trajectory
pairs the minimum separation between which is $\rho\leqslant$ 10 pc.\normalsize
\label{fig:y24o24-2}
\end{figure}
\twocolumn
\subsubsection*{J1915+1009 and J1909+1102}
\noindent
The results of our
trajectory simulations for the pair of J1915+1009 and
J1909+1102 are presented in Fig. 8. Here, the epoch
of a close encounter of the pulsars corresponds to
$T\sim$ 300000 yr. With the characteristic age of the
younger pulsar J1915+1009 $\tau_{ch}$ = 420000 yr, this is
a good correspondence.

\subsubsection*{J1832–0827 and J1836–1008, J1833–0827
and J1836–1008}
\noindent
In the pairs of J1832–0827 and
J1836–1008, J1833–0827 and J1836–1008 (Figs. 9
and 10), two different secondary components can be
associated with the same pulsar (J1836–1008). Here,
we cannot unambiguously choose one of these associations
based only on our estimates of the encounter
probability. The two presumed companions, J1832–0827 and J1833–0827, are similar in characteristic
age, both have high velocities (as does J1836–1008),
and both are at a small distance from J1836–1008.
Our simulations in both cases yielded similar results,
and we can only point out that both encounters
are nonrandom. However, if we compare the values
of $V(5)$, we will see that these velocities are too
high for the association of J1833–0827 and J1836–
1008.
For the pulsar J1836-1008, the velocity $V(5)$
reaches almost 5000 km s$^{-1}$, a value higher than the
velocities of even the fastest known pulsars. For the
association of J1832-0827 and J1836-1008, $V(5)$
for the same pulsar decreases to $\sim$2000 km s$^{-1}$, a
fairly high value but not so implausible. Recall that
among the pulsars with known parallaxes (i.e., with
directly measured transverse velocities), J1509+5531
has the maximum value, which is about 1000 km s$^{-1}$
(Chatterjee et al. 2009). On this basis, it seems that
the association of J1832-0827 and J1836-1008 is
more natural.

However, on the whole, both possible pairs suggest
a high space velocity of J1836–1008, which is
marginal even for the pulsar velocity distribution in
general. This can be an argument against the possible
association of these pairs. Here, it is hard to be guided
by the characteristic ages of the pulsars (J1836–1008
is the older object in both pairs, and if this is the case,
then it might not receive a kick at binary disruption),
because their initial periods are unknown.

\subsubsection*{J1917+1353 and J1926+1648}
\noindent 
The last pair is
J1917+1353 and J1926+1648 (Fig. 11). The probability
$P$ differs from $P_0$ by more than three orders of
magnitude, suggesting a possible association of the
pulsars. Their characteristic ages are similar, 428000
and 511000 yr, and we cannot say which pulsar is
actually younger. The encounter epoch occurs at $T\sim$
696000 yr, which may correspond to the age of one of
the pulsars during the formation of which the binary
was disrupted.

\subsection*{CONCLUSIONS}
\noindent
Out of the initially selected 16 pulsar pairs, 7 pairs
have a common property: the close encounters of their
components in the past cannot be considered random.
The selected pulsars have enhanced probabilities of
close encounters and, hence, there is reason to believe
that they were gravitationally bound in the past,
i.e., they were members of the same CBS that was
subsequently disrupted.

Such disruption is probably the finale of the evolution
of two initially massive stars — the result of the
second supernova explosion.When the primary (more
massive) component during its evolution overfills its
Roche lobe, mass transfer and outflow from the binary
begins and it can lose the bulk of the hydrogen
envelope by the time of collapse. As a result, the mass
loss by the binary during its explosion can be smaller
than that needed for disruption. In contrast, during
the second explosion, the binary is disrupted in most
cases (Zasov and Postnov 2006).

As was mentioned above, it is virtually impossible
to identify the case of binary disruption after the
first explosion, when the secondary

\onecolumn
\begin{figure}[t]
\centering
\includegraphics[width=12cm]{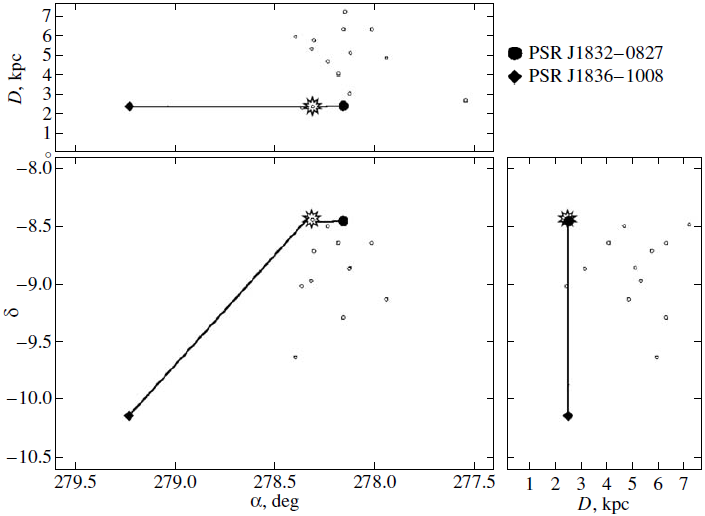}\\
\small {\bf Fig.9:} Same as Fig. 5 for the pulsars J1832–0827 and J1836–1008. The lines correspond to the trajectories at which the
separation found between the pulsars, $\rho_{min}$ = 3.098 pc, is smallest. The points mark the places of the closest approach of
specific trajectory pairs the minimum separation between which is $\rho\leqslant$ 10 pc.\normalsize
\label{fig:y17o7-2}
\end{figure}

\begin{figure}[b]
\centering
\includegraphics[width=12cm]{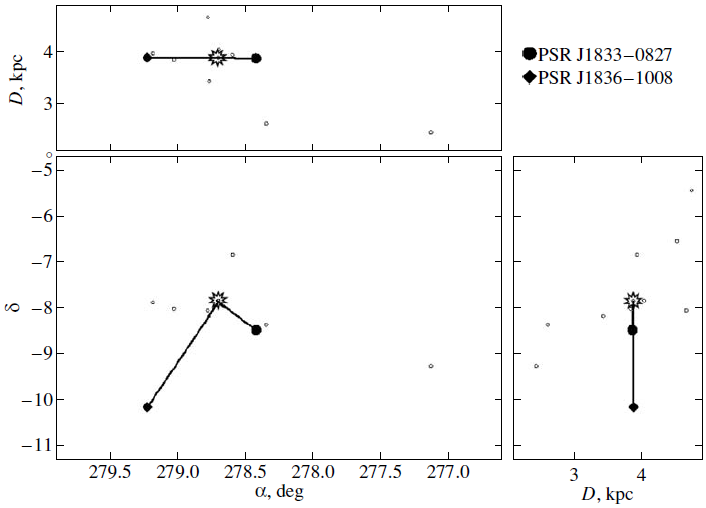}\\
\small {\bf Fig.10:} Same as Fig. 5 for the pulsars J1833–0827 and J1836–1008. The lines represent the trajectories at which the
separation found between the pulsars, $\rho_{min}$ = 1.73 pc, is smallest. The pointsmark the places of the closest approach of specific
trajectory pairs the minimum separation between which is $\rho\leqslant$ 10 pc.\normalsize
\label{fig:y15o7-2}
\end{figure}

\begin{figure}[t]
\centering
\includegraphics[width=12cm]{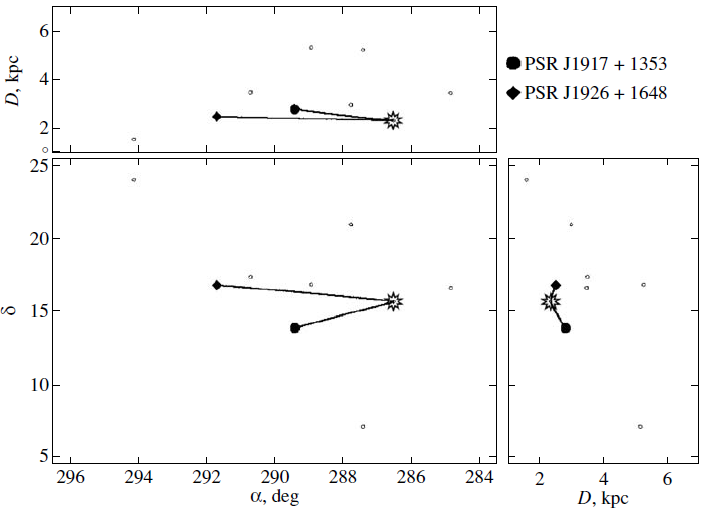}\\
\small {\bf Fig.11:} Same as Fig. 5 for the pulsars J1917+1353 and J1926+1648. The trajectories at which the separation found between
the pulsars, $\rho_{min}$ = 3.399 pc, is smallest are represented by the lines. The points mark the places of the closest approach of
specific trajectory pairs the minimum separation between which is $\rho\leqslant$ 10 pc.\normalsize
\label{fig:y25o1-2}
\end{figure}

\begin{figure}[b]
\centering
\includegraphics[width=12cm]{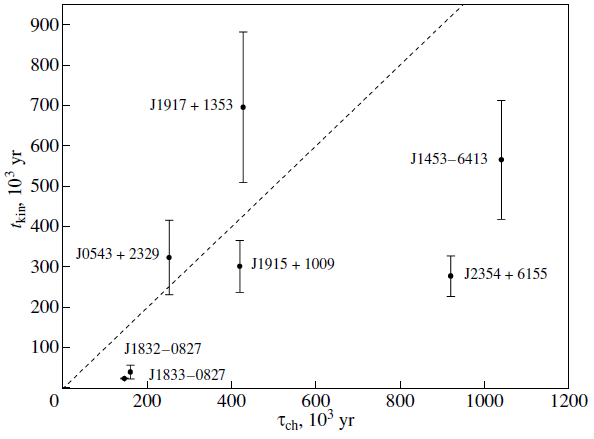}\\
\small {\bf Fig.12:} Relationship between the characteristic ($\tau_{ch}$) and kinematic ($t_{kin}$) ages of the younger pulsars in the pairs. The errors
correspond to the 50\% confidence intervals. The dashed line corresponds to $t_{kin}=\tau_{ch}$.\normalsize
\label{fig:t}
\end{figure}
\twocolumn
\noindent
component, being
now an isolated one, evolves independently from
the primary one. Indeed, after the CBS disruption
through the collapse of the primary component, the
secondary star will recede significantly from the binary
position before its explosion (in $10^5-10^6$ yr). An
additional factor that breaks the connection between
the present-day trajectory of the second pulsar and
the pair coordinates can be an asymmetry in the
explosion of its progenitor. At the same time, our
simulations of trajectories is aimed at detecting the
encounters of pulsars only for a “straight” extension
of the trajectories into the past.

Thus, the encounters
of pulsars we detected probably mark the CBSs that
were disrupted during the second explosion, while
the probability of a purely random close encounter
between pulsars in the past is negligible.

Note that the differences in characteristic ages
of the pulsars in the pairs we selected (except for
the pair of J1917+1353 and J1926+1648) lie within
the range $5$\texttimes$10^5$ - $4$\texttimes$10^6$ yr. These intervals agree
with the time intervals needed for the evolution of the
secondary components to be ended with a supernova
explosion, even if the differences between the characteristic
and actual ages is taken into account.

We analyzed the relations between the characteristic
and ``kinematic'' (determined from the epoch
of closest approach) pulsar ages. Figure 12 shows
the relationship between these parameters. We see
no clear correlation; the kinematic age can be both
older and younger than the characteristic one, but the
age difference itself is small and comparable to the
inaccuracy in $\tau_{ch}$.

The technique used here allows us to determine
whether the kinematic encounters of the selected pulsars
in the past are a consequence of the natural
NS density distribution in the Galaxy or such configurations
are not random. In the former case, the
pair of pulsars under consideration has never constituted
a common binary, although the objects satisfy
the initial selection criteria described above. As an
illustration of this case, let us consider three pulsars:
J1835–1106, J1825–0935, and J1824–1945.
According to the preliminary selection criteria, any of
the first two pulsars is suitable for being associated
with the third. However, neither in the pair of J1835–
1106 and J1824–1945 nor in the pair of J1825–0935
and J1824–1945 can the pulsars be recognized as
being actually associated. Figures 13 and 14 show
the probabilities of detecting the components at a distance
smaller than $\rho$. As we see from the figures, $\rho_0$ for
each pair is very small and the difference between $P$
and $P_0$ is insignificant. Therefore, in this case, the
close encounters of the pulsars can be explained by
purely random factors.

Thus, we detected six pairs of pulsars that were
probably members of disrupted CBSs: J0543+2329
and J0528+2200, J1453-6413 and J1430-6623,
J2354+6155 and J2321+6024, J1915+1009 and
J1909+1102, J1832-0827 and J1836-1008, and
J1917+1353 and J1926+1648. Their further study
by analyzing a larger number of trajectories will
allow us to improve our knowledge about the pulsars
themselves (when the relations between their
kinematic and ``intrinsic'' ($\nu$, $\dot{\nu}$, $\ddot{\nu}$) characteristics are
studied), about the possibility of being members of
common CBSs in the past, and about the disruption
kinematics of such binaries. A reliable identification
of the pulsar pairs will also allow their current radial
velocities and true ages to be estimated.

\subsection*{ACKNOWLEDGMENTS}
\noindent
We are grateful to A.K. Dambis (Sternberg Astronomical
Institute) for a detailed discussion of this
paper and to the referees for helpful remarks. This
work was supported by the “Origin and Evolution of
Stars and Galaxies” Program of the Presidium of the
Russian Academy of Sciences.

\onecolumn
\begin{figure}[t]
\centering
\includegraphics[width=12cm]{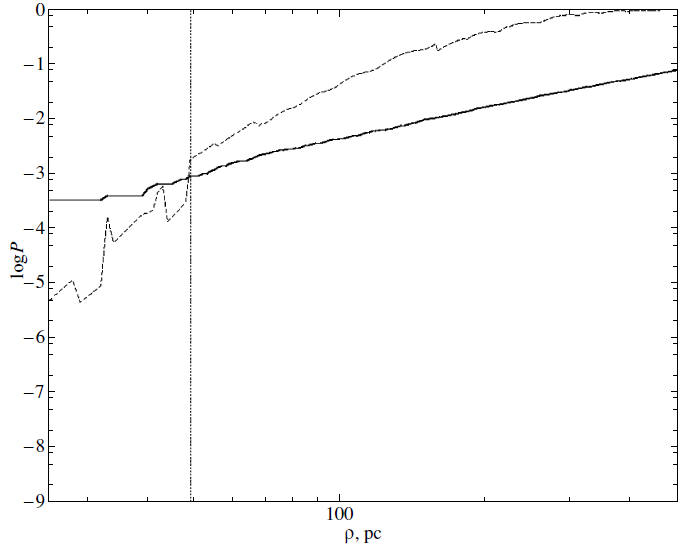}\\
\small {\bf Fig.13:} Pulsars J1835–1106 and J1824–1945. The $\rho$ dependence of $P$ (solid curve) and $P_0$ (dashed curve). The vertical line
indicates $\rho_0$. Here, $\rho_0$ is small, while the probabilities $P$ and $P_0$ differ insignificantly. Therefore, the conclusion that the pulsars
are kinematically associated is invalid.\normalsize
\label{fig:y14o5-1}
\end{figure}

\begin{figure}[b]
\centering
\includegraphics[width=12cm]{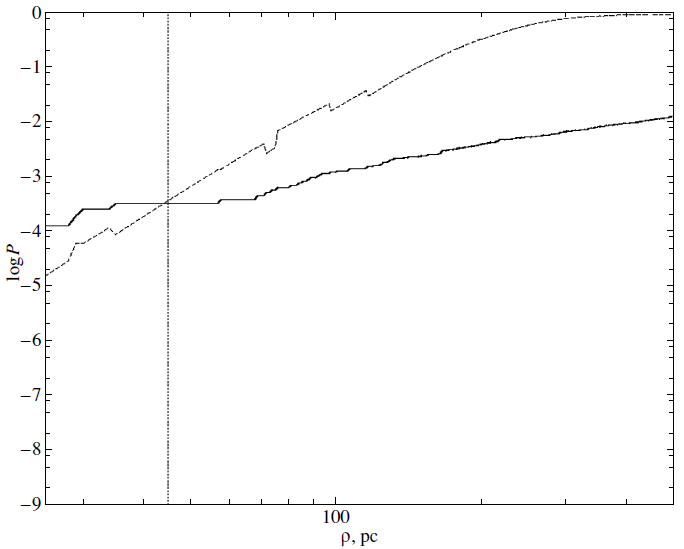}\\
\small {\bf Fig.14:} Same as Fig. 13 for the pulsars J1825–0935 and J1824–1945. Just as the pair of J1835–1106 and J1824–1945, it is
hard to associate the pulsars here.\normalsize
\label{fig:y18o5-1}
\end{figure}
\twocolumn
\small
\subsection*{REFERENCES}
\noindent
[1] Z. Arzoumanian, D.F. Chernoff, and J.M. Cordes,
Astrophys. J. {\bf 568}, 289 (2002).\\\noindent
[2] M. Bailes, Astrophys. J. {\bf 342}, 917 (1989).\\\noindent
[3] A.H. Batten, Ann. Rev. Astron. Astrophys. {\bf 5}, 25
(1967).\\\noindent
[4] G.M. Beskin and S.V. Karpov, Astron. Astrophys.
{\bf 440}, 223 (2005).\\\noindent
[5] G.M. Beskin, V.N. Komarova, S.I. Neizvestny, et al.,
Ex. Astron. {\bf 7}, 413 (1997).\\\noindent
[6] H.A. Bethe and G.E. Brown, Astrophys. J. {\bf 506}, 780
(1998).\\\noindent
[7] G.S. Bisnovatyi-Kogan and B.V. Komberg, Astron.
Zh. {\bf 51}, 373 (1974) [Sov. Astron. 18, 217 (1974)].\\\noindent
[8] A. Blaauw, Bull. Astron. Inst. Netherlands 15, 265
(1961).\\\noindent
[9] V.V. Bobylev, Pis’ma Astron. Zh. {\bf 34}, 757 (2008)
[Astron. Lett. 34, 686 (2008)].\\\noindent
[10] J.A.R. Caldwell and J.P. Ostriker,Astrophys. J. {\bf 251},
61 (1981).\\\noindent
[11] R.G. Carlberg and K.A. Innanen, Astron. J. {\bf 94}, 666
(1987).\\\noindent
[12] S. Chatterjee, W.F. Brisken, W.H.T. Vlemmings,
et al., arXiv:astro-ph/0901.1436v1 (2009).\\\noindent
[13] J.M. Cordes and T.J.W. Lazio, arXiv:astroph/
0207156 (2002).\\\noindent
[14] J.M. Cordes and D.F. Chernoff, Astrophys. J. {\bf 505},
315 (1998).\\\noindent
[15] R.J. Dewey and J.M. Cordes, Astrophys. J. {\bf 321}, 780
(1987).\\\noindent
[16] A. Duquennoy and M. Mayor, Astron. Astrophys.
{\bf 248}, 485 (1991).\\\noindent
[17] C.-A. Faucher-Gigu\`ere and V.M. Kaspi, Astrophys.
J. {\bf 643}, 332 (2006).\\\noindent
[18] B.M. Gaensler and D.A. Frail, Nature {\bf 406}, 158
(2000).\\\noindent
[19] J.R.I. Gott, J.E. Gunn, and J.P. Ostriker, Astrophys.
J. {\bf 160}, L91 (1970).\\\noindent
[20] V.V. Gvaramadze, Astron. Astrophys. {\bf 470}, L9 (2007).\\\noindent
[21] V.V. Gvaramadze, A. Gualandris, and S. Portegies
Zwart,Mon. Not. R. Astron. Soc. {\bf 385}, 929 (2008).\\\noindent
[22] V.V. Gvaramadze, A. Gualandris, and S. Portegies
Zwart, Mon. Not. R. Astron. Soc. {\bf 396}, 570 (2009) \\\noindent
[23] J.L. Halbwachs, M. Mayor, S. Udry, et al., Astron.
Astrophys. {\bf 397}, 159 (2003).\\\noindent
[24] U. Heber, H. Edelmann, R. Napiwotzki, et al., Astron.
Astrophys. {\bf 483}, L21 (2008).\\\noindent
[25] D.J. Helfand and E. Tademaru, Astrophys. J. {\bf 216},
842 (1977).\\\noindent
[26] G. Hobbs, D.R. Lorimer, A.G. Lyne, et al., Mon.Not.
R. Astron. Soc. {\bf 360}, 974 (2005).\\\noindent
[27] I. Iben and A.V. Tutukov, Astrophys. J. {\bf 456}, 738
(1996).\\\noindent
[28] P.D. Kiel and J.R. Hurley, Mon. Not. R. Astron. Soc.
{\bf 395}, 2326 (2009).\\\noindent
[29] K. Kuijken and G. Gilmore, Mon. Not. R. Astron. Soc.
{\bf 239}, 571 (1989).\\\noindent
[30] R.N. Manchester, G.B. Hobbs, A. Teoh, et al., Astron.
J. {\bf 129}, 1993 (2005).\\\noindent
[31] A.M. Mel’nik and Yu.N. Efremov, Pis’maAstron. Zh.
{\bf 21}, 13 (1995) [Astron. Lett. 21, 10 (1995)].\\\noindent
[32] D. Mihalas and J. Binney, {\it Galactic Astronomy:
Structure and Kinematics} (Freeman, San-
Francisco, 1981).\\\noindent
[33] M.P. Muno, J.S. Clark, P.A. Crowther, et al., Astrophys.
J. {\bf 636}, L41 (2006).\\\noindent
[34] B. Paczynski, Astrophys. J. {\bf 348}, 485 (1990).\\\noindent
[35] K.A. Postnov and L.R. Yungelson, Liv. Rev. Relativ.
{\bf 9}, 6 (2006).\\\noindent
[36] A. Poveda, J. Ruiz and C. Allen, BOTT {\bf 4}, 86 (1967).\\\noindent
[37] M.E. Prokhorov and S.B. Popov, Pis’ma Astron. Zh.
{\bf 28}, 609 (2002) [Astron. Lett. 28, 536 (2002)].\\\noindent
[38] N. Przybilla, M.F. Nieva, U. Heber, et al., Astron.
Astrophys. {\bf 480}, L37 (2008).\\\noindent
[39] I.S. Shklovsky, Astron. Zh. {\bf 46}, 715 (1970) [Sov.
Astron. 13, (1970)].\\\noindent
[40] J.H. Taylor and J.M. Cordes, Astrophys. J. {\bf 411}, 674
(1993).\\\noindent
[41] W.-W. Tian and D. Leahy, PABei {\bf 22}, 308 (2004).\\\noindent
[42] W.H.T. Vlemmings, J.M. Cordes, and S. Chatterjee,
Astrophys. J. {\bf 610}, 402 (2004).\\\noindent
[43] G.A.E. Wright, Nature {\bf 277}, 363 (1979).\\\noindent
[44] I. Yusifov and I. K\"{u}c\"{u}k, Astron. Astrophys. {\bf 422}, 545
(2004).\\\noindent
[45] A.V. Zasov and K.A. Postnov, {\it General Astrophysics}
(Vek 2, Fryazino, 2006) [in Russian].

\hspace{3.5cm} {\it Translated by V. Astakhov}

\end{document}